\documentclass[reprint,amsmath,amssymb,aps,prb]{revtex4-2}
\usepackage{xcolor}
\usepackage{graphicx}
\usepackage{caption}
\usepackage{subcaption}
\usepackage{dcolumn}
\usepackage{bm}
\usepackage{appendix}

\usepackage{ulem}

\newcommand{\vtwo}[1]{{#1}}

\begin{document}

\title{Full configuration interaction quantum Monte Carlo for coupled electron--boson systems and infinite spaces}
\author{Robert J. Anderson}
\affiliation{Department of Physics, King’s College London, Strand, London WC2R 2LS, United Kingdom}%

\author{Charles J.~C. Scott}
\affiliation{Department of Physics, King’s College London, Strand, London WC2R 2LS, United Kingdom}%

\author{George H. Booth}
\email{george.booth@kcl.ac.uk}
\affiliation{Department of Physics, King’s College London, Strand, London WC2R 2LS, United Kingdom}%
\date{\today}

\begin{abstract}
We extend the scope of full configuration interaction quantum Monte Carlo (FCIQMC) to be applied to coupled fermion-boson hamiltonians, alleviating the {\it a priori} truncation in boson occupation which is necessary for many other wave function based approaches to be tractable. Detailing the required algorithmic changes for efficient excitation generation, we apply FCIQMC in two contrasting settings. The first is a sign-problem-free Hubbard--Holstein model of local electron-phonon interactions, where we show that with care to control for population bias via importance sampling and/or reweighting, the method can achieve unbiased energies extrapolated to the thermodynamic limit, without suffering additional computational overheads from relaxing boson occupation constraints. Secondly, we apply the method as a ``solver'' within a quantum embedding scheme which maps electronic systems to local electron-boson auxiliary models, with the bosons representing coupling to long-range plasmonic-like fluctuations. We are able to sample these general electron-boson hamiltonians with ease despite a formal sign problem, including a faithful reconstruction of converged reduced density matrices of the system. 
\end{abstract}
\maketitle

\section{Introduction}
Quantum systems of interacting electrons explicitly coupled to bosonic degrees of freedom are ubiquitous in nature. Most notably, electrons can couple to nuclear vibrations, characterized by collective phononic quasiparticles which obey bosonic statistics \cite{Bernardi2016, Giustino2017}. The explicit effect of these lattice vibrations on the electronic structure is neglected in Born-Oppenheimer approximations, but strong coupling between the electrons and these phonons can substantially renormalize the effective interactions in the system, and lead to not just quantitative changing of electronic expectation values, but also qualitative new physics emerging from the system \cite{RevModPhys.82.3045, Froehlich1954, HOLSTEIN1959}. As a famous example, the coupling between electrons and phonons in this way can renormalize interactions of certain solid state systems to such a degree that it can provide the mechanism for the formation of Cooper pairs and superconducting order \cite{Bardeen1973}.

Phonons are not the only bosonic species of physical relevance with which electrons can interact. The emerging field of tailored electronic structure via optical cavities exposes an exciting experimental approach to designing interactions between electrons, and inducing new decoherence pathways for electronic systems \cite{vonderLinden1995, C8SC01043A, Cohen2010}. In this, the photonic field is quantized via quantum electrodynamics, with these bosonic cavity modes explicitly coupled to the electronic degrees of freedom in the dipole limit to form mixed ``polaritonic'' quasiparticles \cite{Ruggenthaler2018, PhysRevLett.115.093001,  PhysRevResearch.2.023262, Rokaj_2018, PhysRevLett.122.193603}. These correlated systems can be found to engage in novel reaction pathways and even room-temperature superfluidity \cite{Lerario2017, Stranius2018}.

Beyond coupling to these ``external'' bosonic fields, collective composite quasiparticles of electrons themselves can manifest with emergent (quasi-)bosonic statistics. These encompass plasmons, magnons and other bosonic-like collective excitations, which can mix strongly with other electronic degrees of freedom \cite{doi:10.1142/S0217979217400070, PhysRevA.90.012508, doi:10.1021/acsphotonics.9b00768,  PhysRevLett.115.093001, PhysRevB.105.014433, W_lfle_2018}. Model Hamiltonians can be constructed whereby the physics of these long-range excitations is included via coupling of a (perhaps local) set of correlated electronic degrees of freedom to explicit bosons describing these physical collective modes. This ``bath'' of bosons can induce novel entanglement, decoherence and screening of the electronic interactions due to the coupling to the electrons \cite{RevModPhys.90.025003, PhysRevLett.62.961, PhysRevB.96.235149, PhysRevB.104.245114}. In this way, the bosons can represent the physical process of a large number of external {\it electronic} degrees of freedom and their effect on a chosen subspace.

However, we are limited in available tools for simulation of these interacting mixed-species quantum systems, which is exceptionally challenging when the instantaneous correlations are fully included, beyond a density functional or Green's function framework \cite{PhysRevB.78.081406, PhysRevLett.121.113002, PhysRevB.93.155102, PhysRevB.84.155104, PhysRevX.3.021011, PhysRevLett.112.215501, PhysRevB.91.155109, PhysRevB.93.100301}. From a wave function perspective, the boson number is generally not conserved in the hamiltonian, leading to a formally infinite Hilbert space of bosonic occupation for any single mode. To ensure wave function approaches are still tractable, it is common to introduce a threshold on the occupancy of any one of these modes, truncating the high-energy parts of the bosonic Hilbert space to ensure a return to a finite-dimensional problem. This approach has been used to develop exact diagonalization \cite{PhysRevB.67.224504, PhysRevB.96.235149}, density matrix renormalization group (DMRG) \cite{PhysRevB.57.6376, doi:10.1021/acs.jctc.7b00329}, coupled-cluster \cite{PhysRevResearch.2.023262, doi:10.1063/5.0033132, PhysRevX.10.041043} and other wave function approaches adapted for coupled electron-boson systems \cite{PhysRevResearch.2.043258}. However, ensuring convergence with respect to these truncations, especially with strongly coupled and/or low-energy boson modes with high average occupation, can still rapidly lead to intractable calculations compared to their fully fermionic counterparts. 

To avoid this truncation, other approaches look to either stochastically sample over all bosonic occupations \cite{Costa2020, assaad, PhysRevB.105.165129}, parameterize an ansatz over the bosonic sector (e.g. as a coherent Gaussian state) \cite{PhysRevB.60.1633, PhysRevB.96.205145, SHI2018245}, or implicitly renormalize over all boson occupations. This latter approach can be achieved in a wave function context within an alternative coupled-cluster formalism, whereby more highly occupied bosonic modes are formed from the product of low bosonic occupation implicitly through the exponential ansatz \cite{doi:10.1063/1.1637578, doi:10.1063/1.4931472, doi:10.1063/1.1637579, doi:10.1063/5.0033132}. Monte Carlo methods on the other hand look to stochastically sample from this space without truncation, but while they represent the {\it de facto} approach for pure bosonic systems, are substantially limited in their applicability to coupled fermion-boson systems by the manifestation of the fermion sign problem in its various guises, which can limit their efficacy \cite{assaad}.

In this work, we look to adapt the modern approach to full configuration interaction quantum Monte Carlo (FCIQMC) to tackle mixed fermion-boson models. This stochastic approach describes an ensemble of signed ``walkers'' which reside on many-body configurations, representing a coarse-grained snapshot of the underlying wave function of interest \cite{BoothThomAlavi2009,initiator-fciqmc,shepherd2012,Booth2013,Overy2014,Shepherd2014,Blunt2015b,Blunt2015_2,BoothF12,
Thomas2015,Thomas2015_2,LiManni2016,Kersten2016,BoothSpectra,Blunt2017,Bogdanov2018,Blunt2018,Blunt2018_2,
guther2018,Samanta2018,LiManni2018,Luo2018,LiManni2019,Dobrautz2019,limanni2019a,Anderson2020,adaptive_shift}. By evolving the walkers according to simple rules underpinned by a master equation derived from the imaginary-time Schr{\"o}dinger equation, expectation values can be averaged over this evolution, providing systematically improvable estimates of correlated observables. Explicit annihilation events between walkers of different sign, as well as ``initiator'' approximations can overcome the fermion sign problem in this context and enable long-time stable dynamics \cite{doi:10.1063/1.3681396, doi:10.1063/1.3302277, doi:10.1063/1.3525712}. This efficient sampling allows for system sizes (and Hilbert spaces) to be considered well beyond the size that can be treated with formally exact approaches, and is emerging as a complementary method to DMRG in fermionic systems. 

The FCIQMC method has been used with success previously both for a wide array of fermionic models \cite{PhysRevB.105.195123, PhysRevB.99.075119, PhysRevB.91.045139, PhysRevB.85.081103}, as well as applied to bosons \cite{PhysRevB.105.235144, condmat7010015}, however has not to date been applied to mixed-species systems. In particular, we believe that an application of FCIQMC to coupled electron-boson systems is particularly appealing, since it is not expected to require an explicit truncation of the bosonic mode occupation, allowing a sampling over all relevant bosonic degrees of freedom. An intrinsic strength of the FCIQMC method is to efficiently seek out relevant parts of the configuration space, and to exploit any natural emergent sparsity in the wave function representation present in low-energy states. This will be of particular importance in the formally infinite Hilbert space of these challenging mixed-species systems, and we expect to provide an important tool for their efficient simulation.

In section \ref{sec:theory} we detail the specific form of the hamiltonians used in this work, while in section \ref{sec:fciqmcadaptations} we describe the key changes required in the FCIQMC algorithm for their efficient simulation, in particular focusing on the excitation generation for general electron-boson hamiltonians. This has been implemented in our FCIQMC package, {\tt M7} \cite{M7}, which extends all modern aspects of FCIQMC functionality, including excited states and semi-stochastic adaptations to these mixed-species models. In Sec.~\ref{sec:HHmodel} we move on to an application of electron-boson FCIQMC to the Hubbard--Holstein model, a paradigmatic model where strongly interacting electrons and local phonon coupling on a lattice gives rise to a number of electronically and phononically driven phase transitions. The approach will be compared to deterministic exact diagonalization results with an explicitly truncated bosonic occupation. Finally, in sec.~\ref{sec:embeddedFCIQMC} we move towards a more general implementation, where electron-boson FCIQMC is used to solve a model of electrons coupled to long-range bosonic quasi-particles. This model requires more general bosonic coupling to arbitrary density fluctuations in the electronic states, as well as general (four-point) electronic interactions amongst the electrons. Furthermore, we demonstrate convergence of density matrix sampling in this general context, as well as algorithmic changes required for excitation generation towards an {\it ab initio} setting.

\section{Electron-boson hamiltonians} \label{sec:theory}
The general form of the second-quantised Hamiltonian considered in this work can be written as
\begin{align}
\label{eq:ham}
\hat{H} &= \hat{H}_\mathrm{elec} + \sum_m \omega_m \hat{a}^\dagger_m \hat{a}_m  \nonumber\\
 &+\sum_{mpq}
\sum_{\sigma \in \{ \uparrow, \downarrow\}}
 (V_{m p_\sigma q_\sigma}\hat{c}^\dagger_{p_\sigma}\hat{c}_{q_\sigma}\hat{a}_m + g_m\hat{a}_m) + \mathrm{h.c.}
\end{align}
where $\hat{c}$ symbolises the fermionic annihilation operator, and $\hat{H}_\mathrm{elec}$ contains the standard one- and two-electron (interaction) terms over the single fermion states (labelled $p, q$) and spin labels $\sigma, \tau$:
\begin{align}
    \hat{H}_\mathrm{elec} &= 
    \sum_{pq}\sum_{\sigma \in \{ \uparrow, \downarrow\}} h_{p_\sigma q_\sigma} \hat{c}^\dagger_{p_\sigma}\hat{c}_{q_\sigma}\nonumber \\&+
    \sum_{pqrs}\sum_{\sigma\tau \in \{ \uparrow, \downarrow\}} g_{p_\sigma q_\tau r_\sigma s_\tau} \hat{c}^\dagger_{p_\sigma}\hat{c}^\dagger_{q_\tau}\hat{c}_{s_\tau}\hat{c}_{r_\sigma}
\end{align}
The bosonic annihilation operator is denoted $\hat{a}$, with indices $m$ extending over all modes. These bosons can represent phononic, photonic or other arbitrary bosons, which in this work will interact in a linear fashion with the quantum fluctuations in the fermionic charge density (although these can in principle couple to e.g. spin or pairing fluctuations or other higher-order electronic operators). The effect of this explicit coupling on the electronic structure is a dynamic or energy-dependent change the effective interactions between the electrons, mediated by these bosons, to screen the electrons at different length or energy scales in the system.

A simplified paradigmatic model system which captures these effects as a special case of Eq. \ref{eq:ham}, is the Hubbard--Holstein model \cite{von_der_Linden1995, PhysRevB.52.4806, PhysRevB.31.6022, PhysRevB.41.11557, PhysRevB.54.2410, PhysRevB.57.5051, PhysRevLett.95.096401, KOLLER2005795, PhysRevB.76.155114, PhysRevB.88.125126, doi:10.1021/acs.jctc.8b01116, PhysRevB.94.085115}, as
\begin{equation}
\label{eq:hh_ham}
\hat{H} = \hat{H}_\mathrm{Hubbard} + 
\sum_{m=1}^{L} g\hat{n}_m (\hat{a}_m + \hat{a}_m^{\dagger}) + \sum_{m=1}^L \omega_0 \hat{a}^\dagger_m \hat{a}_m
\end{equation}
where $\hat{n}_m\equiv \hat{n}_{m_\uparrow} + \hat{n}_{m_\downarrow} \equiv \hat{c}^\dagger_{m_\uparrow}\hat{c}_{m_\uparrow}+\hat{c}^\dagger_{m_\downarrow}\hat{c}_{m_\downarrow}$ is the fermion number operator, and the electronic $\hat{H}_\mathrm{Hubbard}$ is parametrised as
\begin{align}
    \hat{H}_\mathrm{Hubbard} &= -t \sum_{\langle i,j \rangle} \sum_{\sigma \in \{ \uparrow, \downarrow\} } 
    \left(
    \hat{c}^{\dagger}_{j_\sigma} \hat{c}_{i_\sigma} + \hat{c}^{\dagger}_{i_\sigma} \hat{c}_{j_\sigma}
    \right) \nonumber \\ &+ U \sum_{i=1}^L \hat{n}_{i_\uparrow} \hat{n}_{i_\downarrow} 
\end{align}
by the usual hopping between nearest neighbours ($\langle i,j \rangle$) and on-site repulsion term.
In this way, the sites of the Hubbard--Holstein model are each coupled with strength $g$ to a local boson mode of frequency $\omega_0$. The effect of these local boson modes, which can drive effective electronic pairing between the fermions, compete with both the on-site electronic repulsion and single-particle kinetic energy terms. It is a prototypical model for phonon-driven emergence of charge density waves in correlated materials and Peierls instabilities, with metallic, Mott insulating and charge density modulated phases present depending on the dominant terms, representing an important and unsolved model system.

\section{Adaptation of FCIQMC for electron-boson systems} \label{sec:fciqmcadaptations}

Due to the coupling with bosonic degrees of freedom, the many-body Hilbert space in which the FCIQMC walker population must evolve is one formed by the product of fermionic and bosonic configurations. 
While the basic ingredients of the algorithm are largely the same, the necessities of generating boson-coupled excitations and accommodating bosonic occupation in the many-body basis are significant departures from the assumptions of a determinant space implementation.
While the determinant is stored as a bit string, the bosonic part of the configuration must be stored as an integer array allowing arbitrary occupancy \cite{PhysRevB.105.235144}. Since the Hamiltonian in Eq. \ref{eq:ham} does not conserve boson number, there is in theory no limit to the number of bosons that can be occupied in the CI space basis. Practically speaking, high boson occupation is energetically penalised by the positive frequency $\omega_n$ of the modes, with the expected distribution of bosonic occupation following a Poissonian distribution. This implies an exponential accuracy in representation of the wave function with increasing truncation of the maximum boson occupation. In practice, we store the boson occupations in the FCIQMC results of this paper as an array of unsigned characters, giving a maximum cutoff occupation of 255 bosons per mode (though we can control this in order to enable direct comparison to approaches with lower occupation truncations). An occupancy of 255 is well beyond any reasonable occupation visited during the stochastic dynamics on any practical timescale, and so this can be taken to be effectively infinite within the sampling approximation.

\subsection{Recap of FCIQMC algorithm}
FCIQMC \cite{BoothThomAlavi2009, linear-scaling-fciqmc} is a projector monte carlo method which represents the many-body wavefunction as a population of signed walkers at each projective iteration. The average number of walkers on each many-body basis function will - given a total number of walkers sufficient to resolve any present sign problem - converge to the exact FCI coefficient. Thus, expectation values with respect to the ground state wavefunction can be extracted by averaging their values over a number of walker distributions generated by iterative application of a stochastically-realised integrator.

Usually, this is based on the linear order truncation of the imaginary time evolution operator, giving 
\begin{equation}
\label{eq:proj_udpate}
\Delta C_\mathbf{i} = - \Delta\tau\sum_\mathbf{j\neq i} H_\mathbf{ij} C_\mathbf{j} - \Delta\tau (H_\mathbf{ii} - E_S)C_\mathbf{i} 
\end{equation}
as the average update of the coefficients.
Expressed in the first term are the off-diagonal {\it spawning} contributions. These are stochastised by making on average $|C_\mathbf{j}|$ attempts to draw basis functions $\mathbf{i}$ with non-zero matrix element $H_\mathbf{ij}$ in a process referred to as {\it excitation generation}.
The cost of the calculation therefore scales linearly in the 1-norm of the amplitudes, or number of walkers
$N_W\equiv \sum_\mathbf{i} |C_\mathbf{i}|$.

Production scale calculations with FCIQMC codes require parallelisation, with each basis function assigned a rank index within a message passing communicator. Spawned contributions are sent via this interface to the process on which the sampled $C_\mathbf{i}$ would be accumulated. The incoming spawns are combined with each other and any existing walker population $C_\mathbf{i}$ in the {\it annihilation} procedure. Sufficient incidence of annihilation between positive and negative walkers is understood to be the crucial factor in suppressing the growth of the unphysical {\it stoquastic} signal, and thus overcoming the sign problem \cite{doi:10.1063/1.3681396, Shepherd2014}. In problems with a severe sign problem, the required number of walkers can approach the FCI Hilbert space dimension - nullifying the advantages of the sparse representation afforded by the FCIQMC approach. In response to this, the {\it initiator} approximation was introduced, which imposes a dynamic truncation of the Hamiltonian prohibiting spawns from low-weight walkers in exchange for an approximate, but systematically-improvable sampling of a physical ground state eigenvector. The initiator approximation is most successful when the wavefunction is localised in orbital space, as is often the case in weak-to-moderately correlated MO-based {\it ab initio} or reciprocal space lattice model systems, wherein systematic errors can be reduced to within acceptable bounds well before the unmodified FCIQMC would be able to tractably sample the physical ground state.

The second term in Eq. \ref{eq:proj_udpate} is a diagonal scaling term which includes the {\it shift} $E_S$.
By modulating the shift, the total walker creation due to spawning/annihilation can be matched by the number removed via this diagonal {\it death} process, maintaining an approximately constant $N_W$ and providing in $E_S$ an estimation of the energy expectation value associated with the sampled wavefunction.

\subsection{Generalized Excitation Generation} \label{sec:Excitgen}

Excitation generation constitutes a core component of FCIQMC, whereby individual off-diagonal terms in the hamiltonian are selected according to a normalized probability distribution for a given configuration, providing another ``connected'' configuration in the full Hilbert space \cite{doi:10.1021/acs.jctc.8b00844, doi:10.1080/00268976.2013.877165, doi:10.1021/acs.jctc.5b01170, doi:10.1063/5.0005754}. For maximum efficiency, these terms should be selected with a probability approximately proportional to the absolute magnitude of the hamiltonian matrix element connecting the configurations. Since this will in general depend on the configuration in question, this can not be efficiently achieved in practice, and so a number of approaches have been developed, aiming at minimizing the cost of this stochastic selection algorithm, and ensuring as optimal a probability distribution in this sampling as possible.
We initially divide these types of $|\mathbf{i}\rangle \rightarrow |\mathbf{j}\rangle$ excitations into different classes, assigning initial probabilities of these class types to each. These now include both fermionic single and double excitations, as well as the (separate) probabilities for bosonic excitation and de-excitations. The first step involves selecting a class of excitation to generate according to these normalized probabilities, and the probabilities of generation within each class are continually optimized through the calculation in order to minimize multiple walker creation events in a single attempt, and therefore allow the largest timestep possible with stable dynamics.

The algorithm for generating excitations from the electronic part of the hamiltonian proceeds as normal, with single excitations $|\mathbf{i}\rangle \rightarrow \hat{c}^\dagger_{i}\hat{c}_{j}|\mathbf{i}\rangle$ (with $i,j$ distinct) drawn by uniform selection among occupied and unoccupied spin orbitals of the same spin, taking into account any abelian symmetry constraints of these orbitals. Double excitations are drawn as $|\mathbf{i}\rangle \rightarrow \hat{c}^\dagger_{i}\hat{c}^\dagger_{j}\hat{c}_{l}\hat{c}_{k}|\mathbf{i}\rangle$ (with $i,j,k,l$ all distinct) according to the precomputed heat bath algorithm (PCHB), which approximately draws these indices according to the magnitude of the hamiltonian matrix elements \cite{doi:10.1021/acs.jctc.8b00844,
doi:10.1021/acs.jctc.5b01170, doi:10.1063/5.0005754}. 
We focus here on the boson coupling term, which introduces two novel excitation generation cases, namely fermion hopping coupled to boson excitation and de-excitation, as
\begin{equation}
\label{eq:bosonex}
|\mathbf{i}\rangle \rightarrow \hat{c}^\dagger_{i}\hat{c}_{j}\hat{a}^\dagger_n|\mathbf{i}\rangle; \quad
|\mathbf{i}\rangle \rightarrow \hat{c}^\dagger_{i}\hat{c}_{j}\hat{a}_n |\mathbf{i}\rangle
\end{equation}
(with $i,j$ distinct). Boson number-conserving terms in the Hamiltonian are diagonal, contributing only to the death rate of walkers, and so do not need to be considered in the excitation generation.

A stochastically efficient excitation generator for the connections of Eq. \ref{eq:bosonex} is straightforward to achieve via a reapplication of the principles of the heat bath algorithm to effectively weight the generation of these terms. To ensure hermiticity, the coefficients $V_{ijn}$ of the boson excitations are the same as those $V_{jin}$ for the de-excitations, and so a single PCHB sampler suffices for both cases.
When the PCHB sampler is invoked for a boson excitation, a uniform single excitation $|\mathbf{i}\rangle \rightarrow \hat{c}^\dagger_{i\sigma}\hat{c}_{j\sigma}|\mathbf{i}\rangle$ is drawn, with the $(i,j)$ compound index referencing a row in the PCHB table. 
Note that this selection of the $(i,j)$ pair is not invariant to the order of the selection of the two indices, differing from the two-electron integral precomputed table, which uses a triangular mapping to leverage the $i<j$ permutational symmetry of the uniformly picked annihilation operator pair.

A bosonic index is then picked with probability $|V_{ijn}|/\sum_m|V_{ijm}|$, whose values are precomputed in an alias table for efficient selection. This ensures that the these bosonic excitations are drawn from a probability distribution as close as possible to the true hamiltonian magnitude between the states. Unlike the fermionic PCHB implementation, drawn excitations are only discarded as null when mode $n$ is already at the imposed $n_\mathrm{boson}$ cutoff (which never occurs in practical settings where the cutoff is 255). De-excitations are drawn in exactly the same way, except that the indices $i,j$ are exchanged to find the correct row in the sampler table, and are discarded when mode $n$ is unoccupied.

\section{1D Hubbard--Holstein model} \label{sec:HHmodel}

We first test the ability to faithfully sample from the boson spaces in these models via the 1D Hubbard--Holstein model, as given in Eq.~\ref{eq:hh_ham}. Due to the form of the model, the excitation generation is simplified from its generality described in Sec.~\ref{sec:Excitgen}. The electronic hopping is stochastised by simply selecting an occupied spin orbital and then randomly choosing to move left or right, generating a null excitation if the adjacent spin orbital is already occupied. Furthermore, since the model only couples bosonic operators to the electronic density (in the site basis), not the hopping, the parent configuration bit-string is also used for the bosonic de-excitation generation to select a boson mode.
This is efficiently achieved by caching the occupied sites coordinated to a boson mode with non-zero occupation.
An analogous list is not prepared for the boson creation excitation generator, since in general one is interested in the limit in which no practically no cutoff is imposed on mode occupations, and therefore all modes coupled to occupied sites can be excited.

Even though the term in the Hubbard--Holstein model Hamiltonian which introduces coupling between the electronic density and bosonic (de-)excitations has a positive coefficient (therefore inducing both positive and negative walkers into the dynamics), the overall FCIQMC simulation in the site basis of the model is sign-problem-free. This is because these bosonic interactions only connect adjacent boson number sectors of the Hilbert space, and ensures that the sign on any single configuration is constrained to a single sign, precluding annihilation events from this part of the hamiltonian. Walkers that are positive in one boson number sector are sign-coherently connected to negatively-signed walkers in the adjacent sectors. Furthermore, if there is an odd (even) number of electrons in each spin sector with periodic (anti-periodic) boundary conditions, the Hubbard hopping term is also sign-problem-free, while the interactions are diagonal in the basis.
Despite being a sign-problem-free system, the model can still be challenging to solve and is the subject of continued research, making it an ideal test bed for the efficiency and convergence of the FCIQMC algorithm.

\begin{figure}
    \centering
    \includegraphics[width=0.45\textwidth]{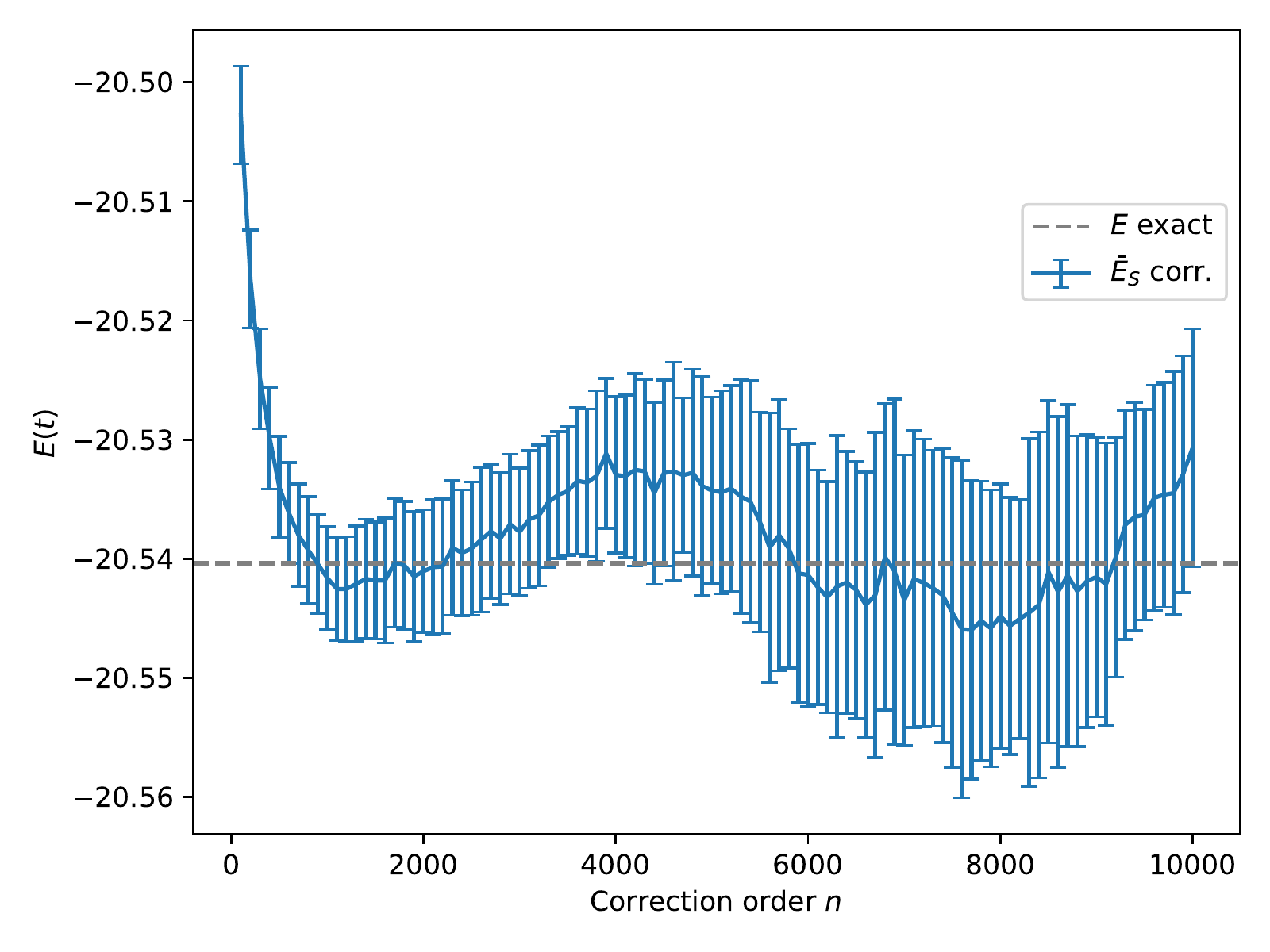}
    \caption{Correlation energies for the Hubbard--Holstein model ($L=8$, $n_\mathrm{elec}=8$, $U=2.0t$, $\omega_0=5.0t$, $g=\sqrt{10.0}t$) with maximum boson occupation per mode of $n_\mathrm{boson}=1$, compared to the exact diagonalisation value. These energy estimates are unbiased for population control bias by the reweighting of shift estimates of the energy. This correction is performed across the time series of shift measurements ($E_S$) and instantaneous $N_W$ for different correction orders ($n$). Consecutive points in the time series from which the corrected energies are computed were 10 Monte Carlo iterations apart, and the shift updates occur with the same frequency and with damping parameter $\gamma=5$. The FCIQMC was performed for $16\times 10^6$ MC cycles with a mean walker number of $\bar{N}_W=188$.}
    \label{fig:correction_order}
\end{figure}

Due to this lack of sign problem, the exact ground state eigenvalue can in principle be estimated from the shift estimator ($E_S$) with an arbitrarily small $N_W$, given a long enough evolution of the walker population and a correction for the population control bias.
It has been shown in recent work \cite{PhysRevB.103.155135} that this correction can be made without alterations to the FCIQMC algorithm, and that given only the $N_W$ and shift time series the systematic bias due to the statistical covariance between the shift and the stochastic wave function can be substantially removed \cite{PhysRevLett.60.1562, PhysRevB.105.235144, PhysRevE.51.3679, doi:10.1063/1.465195}. 
The essence of this approach is to retrospectively ``undo'' the dynamical scaling of the wavefunction due to the shift-dependent death step, so that the covariance between the wavefunction and the shift can be reduced.
The crucial factor to be determined is the optimal span of time series points $n$ on which these ``reweightings'' should be performed so as to effectively remove the bias without introducing excess random error due to the reduction in number of effectively sampled points. We call $n$ the ``correction order'' in Fig.~\ref{fig:correction_order}.

We first test this approach for a small lattice where we can perform exact diagonalization calculations (8 sites). Obtaining the exact diagonalization data also necessitates a severe truncation in the maximum boson occupation, which we limit to a single boson occupancy per mode. The shift-bias corrected FCIQMC data with the same equivalent artificial boson truncation is shown in Fig.~\ref{fig:correction_order}. Note that the figure shows the expected unbiased energy as a function of the length of history of the walker population used in order to control for population bias, not iteration number. This shows similar characteristics to the equivalent plot in Ref.~\onlinecite{PhysRevB.103.155135} for a single calculation. Initially there is convergence towards the exact eigenvalue, then we observe the expected diminishing returns of attempting to further reduce systematic error by elongating the history over which the correction is computed, as the statistical uncertainty in the estimator grows. For this model, a correction order around 2000 appears to be a good compromise. This however serves as a proof-of-principle of the exactness of the implementation, and ability to compute unbiased energies in these (albeit small) systems compared to exact results.

\subsection{Importance Sampling}

The bias in the uncorrected shift estimator for sign-problem-free systems has also been shown to reduce significantly with the introduction of importance sampling, the basic premise of which is to sample not the ground state CI coefficients $C_\mathbf{i}$, but the modified wave function $\tilde{C}_\mathbf{i} = \langle \phi_\mathrm{guide} | \mathbf{i} \rangle C_\mathbf{i}$ where $|\phi_\mathrm{guide}\rangle$ is a ``guiding'' wave function.
Equivalently viewed as a similarity transform of the Hamiltonian, this approach effectively modifies the spawning probabilities such that moves to configurations with larger diagonal matrix element are probabilistically suppressed. 
Expectation values can then be unbiased for this modification to the dynamic. 
\cite{doi:10.1126/science.231.4738.555, PhysRevB.86.075109} Even with a very simple form for the guiding wave function, such as that of the Gutzwiller-like form
\begin{equation}
  |\phi_\mathrm{guide}\rangle = \sum_\mathbf{i} e^{-\alpha H_\mathbf{\mathbf{ii}}} |\mathbf{i}\rangle , \label{eq:Gutz}
\end{equation}
significant improvements have been found in fermionic model systems. We expect this guiding wave function to be even more powerful in the case of boson number non-conserving Hamiltonians where inhibition of spawning to more energetic configurations with highly occupied bosonic modes is an obvious route to achieving a more compressed stochastic representation of the ground state.

\begin{figure}
     \centering
     \begin{subfigure}[b]{0.45\textwidth}
        \centering
        \includegraphics[width=\textwidth]{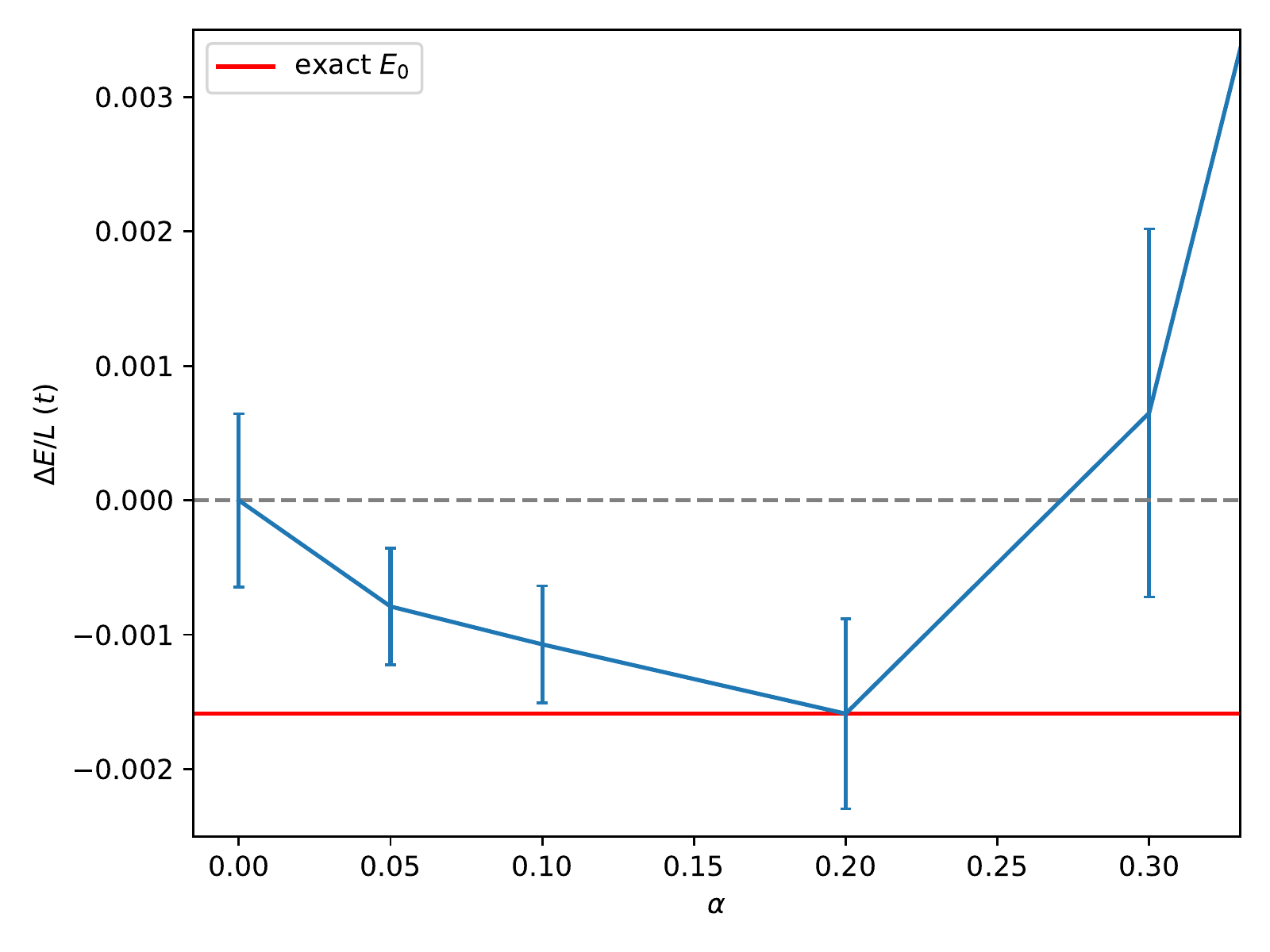}
        \caption{8-site chain with walker number of $N_W = 2\times 10^4$ and $n_\mathrm{boson}=2$, an exactly diagonalizable system.}
        \label{fig:imp_samp_exact}
     \end{subfigure}
     \begin{subfigure}[b]{0.45\textwidth}
        \centering
        \includegraphics[width=\textwidth]{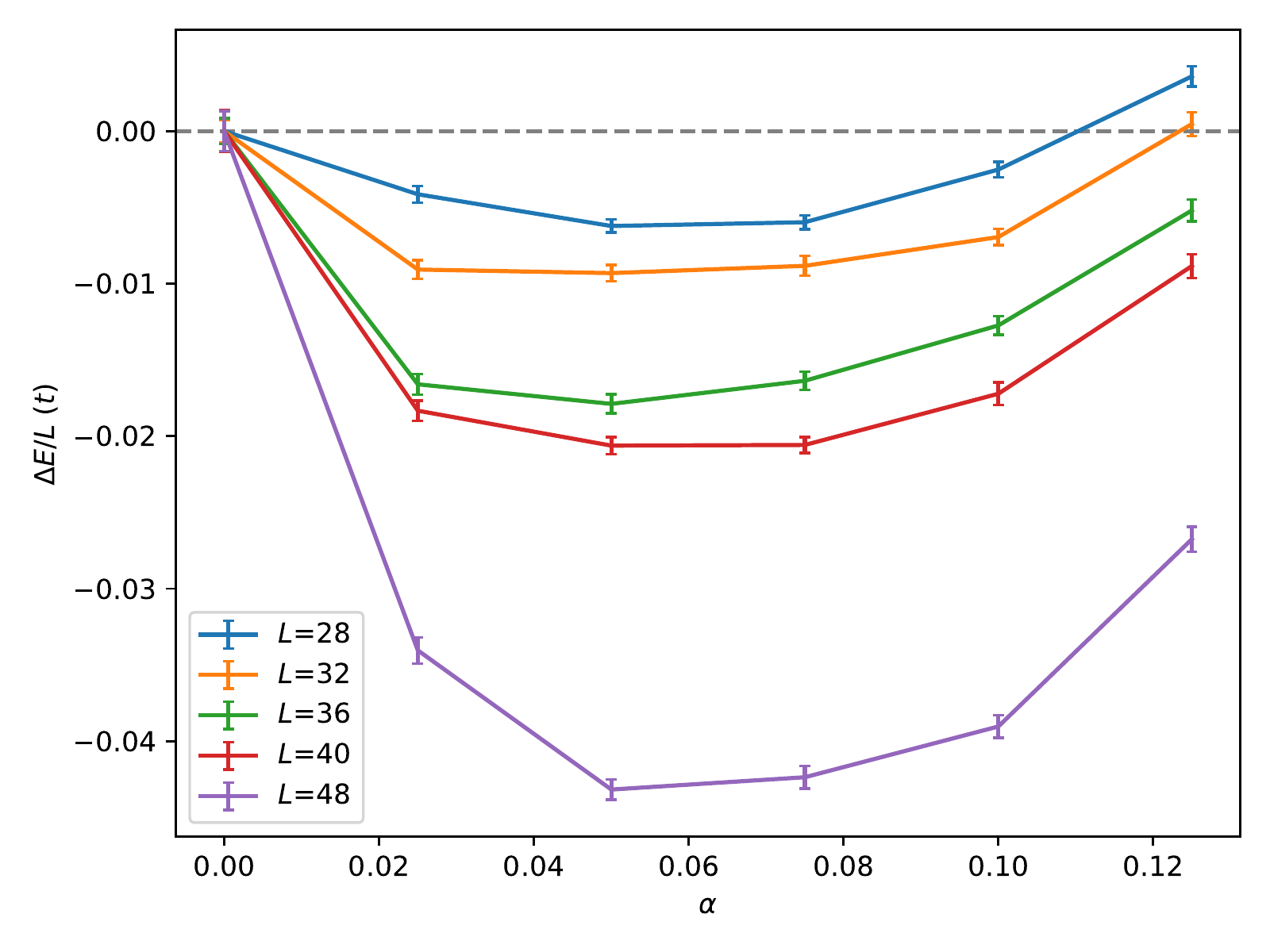}
        \caption{Reduction in population control bias for increasing Hubbard--Holstein chain lengths. Each series was obtained with $n_\mathrm{boson}=255$ and a target $N_W$ of $10^6$.}
        \label{fig:imp_samp_long}
     \end{subfigure}

    \caption{Population control bias amelioration by importance sampling of the Hubbard--Holstein ground state ($U=2.0t$, $\omega_0=5.0t$, $g=\sqrt{10.0}t$) using a guiding wave function of the type defined in Eq.~\ref{eq:Gutz}, with variation in the single adjustable parameter, $\alpha$. All energies are shown relative to the average shift with no importance sampling ($\alpha=0$).}
    \label{fig:imp_samp}
\end{figure}

Figure \ref{fig:imp_samp_exact} illustrates an example application of this adaptation where an $\alpha$ value of $\sim 0.2$ is seen to remove the bias almost completely, even without the explicit reweighting algorithm of Fig.~\ref{fig:correction_order}, though a significant bosonic occupation cutoff is applied ($n_\textrm{boson}=2$) in order to enable comparison to exact diagonalization. 
The plot also shows the expected trend of increasing random error in the shift estimator due to the importance sampling \cite{PhysRevB.103.155135}. Intuitively, this is due to the reciprocal relationship between the importance sampling factors for proposed moves $\mathbf{i}\rightarrow\mathbf{j}$ and $\mathbf{j}\rightarrow\mathbf{i}$. As a move $\mathbf{i}\rightarrow\mathbf{j}$ becomes more suppressed, the variance of the walker population (and true CI coefficient) on $\mathbf{j}$ increases, and the back-spawning magnitudes also increase. This boosted reproduction rate of ill-defined walker occupations propagates more error. Fortunately, it seems that the increases in noise are modest in comparison to the dramatic reduction in the systematic sampling error due to population control bias.


\subsection{Relaxing Bosonic Occupation Constraints}

Having established the ability to practically obtain unbiased ground state energetics via electron-boson FCIQMC in small systems with boson occupation constraints, we now turn to the practicality of relaxing these boson occupation constraints of any mode. This requires the FCIQMC to effectively sample from an infinite Hilbert space of configurations comprising all possible combinations of boson occupations of each mode. We will consider the convergence with respect to relaxing the bosonic threshold on both the (population bias corrected and uncorrected) ground state energies, as well as the convergence of these FCIQMC estimates with respect to the number of walkers.

We start by considering the Mott regime of the 8-site Hubbard--Holstein model, with $U=4t$, $\omega_0=0.5t$ and $g=\sqrt{0.15}t$. In this hamiltonian, the phononic renormalization is not sufficient to overcome the strong Coulomb repulsion, and Mott order remains in the system. While strongly correlated, the relatively weak boson coupling should allow for consideration of the convergence of the energies with number of walkers compared to exact diagonalization as the occupation is increased. Furthermore, we can transform the hamiltonian, by performing a shifting of the boson operators to remove the coupling of the phonons to the {\it static} charge density of the system (which is known exactly by symmetry). This zero phonon mode removal (ZPMR) transformation can leave the exact eigenstates of the full Hamiltonian unchanged (as long as it is accounted for in the expectation values), but can dramatically increase the speed of convergence with respect to average bosonic occupation. This is because the bosons now only couple to charge {\it fluctuations}, rather than the mean-field static component. This allows exact diagonalization to converge faster with boson occupation, and we can compare this to the unconstrained FCIQMC as walker numbers increase. We note that while this transformation of the hamiltonian can in principle also be performed for FCIQMC, it is not in this case, since it would induce a sign problem in the resulting dynamics. The details of this transformation are given in Appendix~\ref{app:zpmr}.

\begin{figure}
    \centering
    \includegraphics[width=0.45\textwidth]{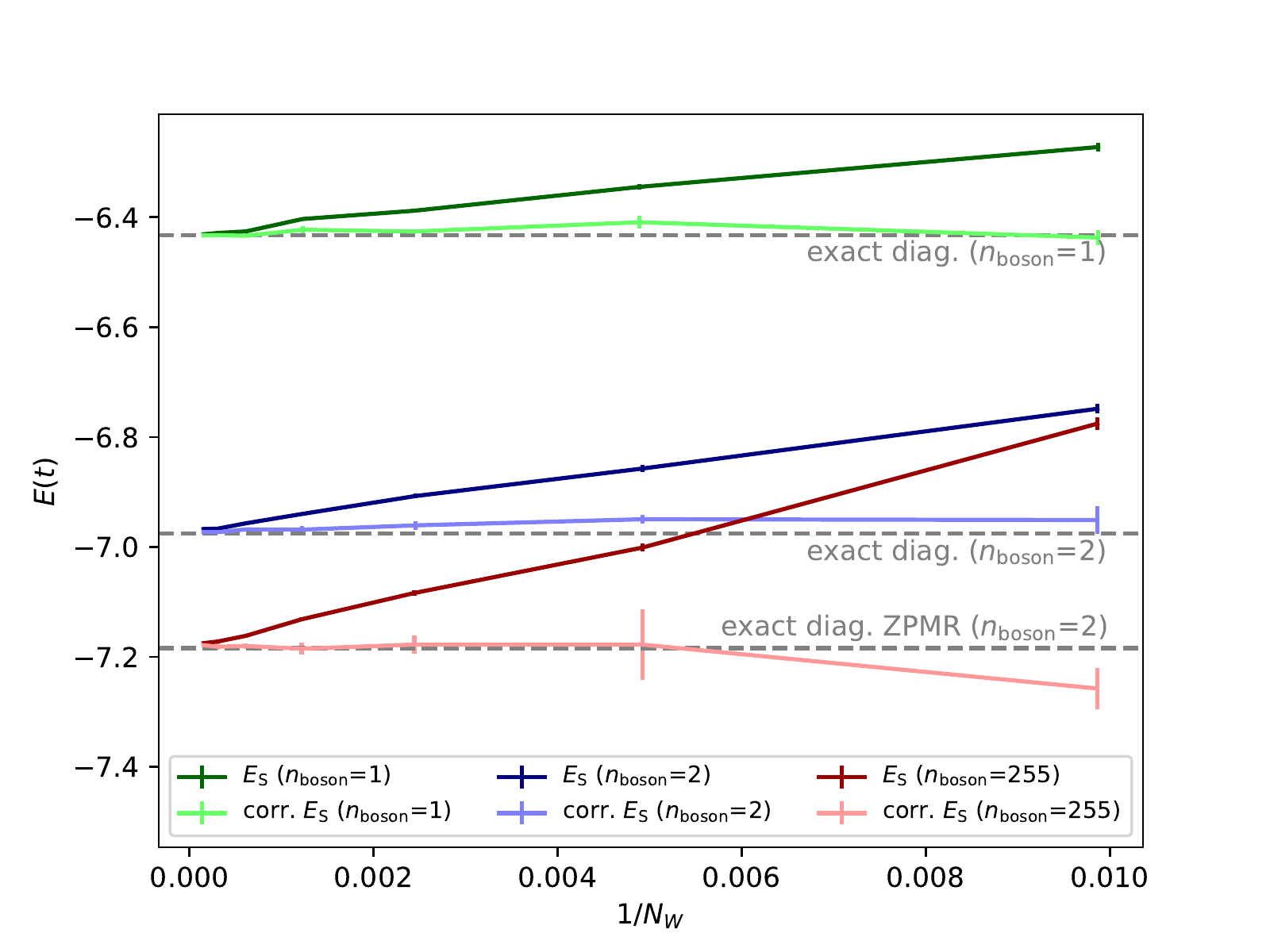}
    \caption{Approximately linear convergence of systematic energy errors with respect to $1/N_W$ in the Mott insulator phase of the Hubbard--Holstein model at half filling with $L=8$, $U=4t$, $\omega_0=0.5t$, and $g=\sqrt{0.15}t$, as well as energies explicitly corrected for this bias showing agreement with exact diagonalization results. We show uncorrected ($E_S$) and explicitly reweighted (corr. $E_S$) average energy (shift) estimators  from the FCIQMC, for the model with $n_\textrm{boson}$ truncated to 1, 2 and untruncated ($n_\textrm{boson}=255$ for practical purposes). These first two results are compared to exact diagonalization in the same Hilbert space, while we compare the untruncated FCIQMC results to exact diagonalization with the shifted bosonic operators and $n_\textrm{boson}=2$, which should be well converged in this regime (see Appendix \ref{app:zpmr}), showing good agreement of the corrected results at all walker numbers.}
    \label{fig:hh_mott}
\end{figure}

Fig.~\ref{fig:hh_mott} shows the convergence of the ground state FCIQMC energy estimates with respect to reciprocal walker number, for both the uncorrected and  {\it a posteriori} corrected shift energy estimates. As has been noted elsewhere (although still somewhat debated), we find a linear convergence of the population control bias with respect to reciprocal walker number. While this error is reduced with the importance sampling of the dynamic via the trial wave function, this doesn't sufficiently remove the bias, and the reweighting approach is still required.

Once this reweighting is applied, the energies at all walker numbers are unbiased, with the exact diagonalization results for a boson occupation truncation of one and two within the statistical errors of the FCIQMC. Relaxing the boson occupation entirely ($n_\textrm{boson}=255$ for technical reasons, but this threshold is never reached) the energies are relaxed by a further $\sim0.2t$. These agree well with the exact diagonalization result in the presence of the ZPMR transformation of the boson operators, despite a boson occupation truncated to $n_\textrm{boson}=2$, which we believe to also be well converged in this regime with weak boson coupling. We find our best FCIQMC result for the model to be $-7.1787(17)t$ at $\sim 7\times 10^3$ walkers, with the ZPMR exact diagonalization result with $n_\textrm{boson}=2$ to be $-7.18408t$.

\begin{figure}[t!]
    \centering
    \begin{subfigure}[b]{0.45\textwidth}
        \centering
        \includegraphics[width=\textwidth]{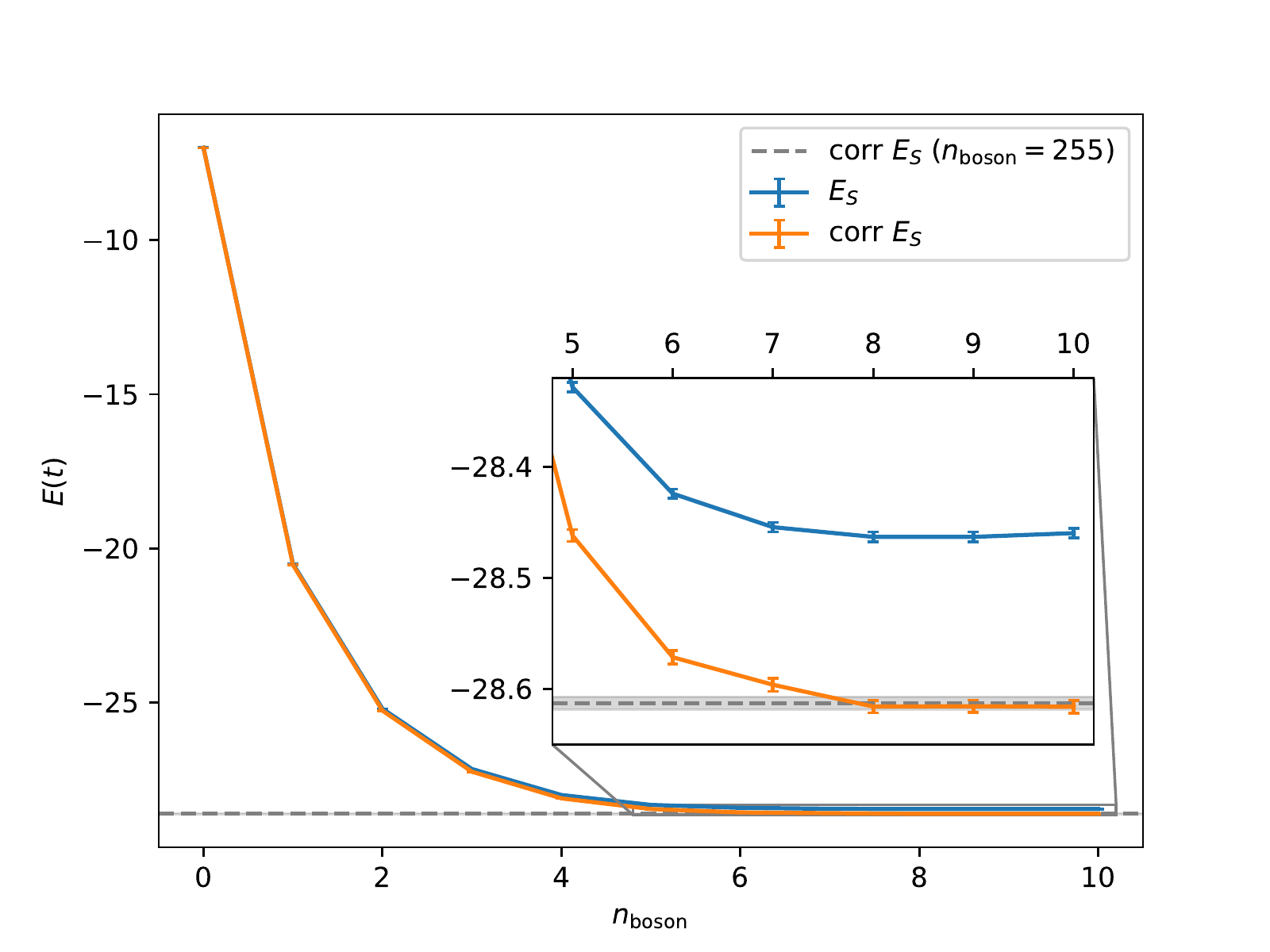}
        \caption{Convergence of the uncorrected and corrected shift energies for the ground state with respect to $n_\mathrm{boson}$ truncation. The shaded area represents the random error in the corrected shift energy with $n_\mathrm{boson}=255$ (in practice, an untruncated bosonic occupation). All means and standard errors were obtained by blocking \cite{doi:10.1063/1.457480} of the shift time series with $\bar{N}_W \sim 800$.
        \label{fig:conv_nboson}}
    \end{subfigure}
    \begin{subfigure}[b]{0.45\textwidth}
        \centering
        \includegraphics[width=\textwidth]{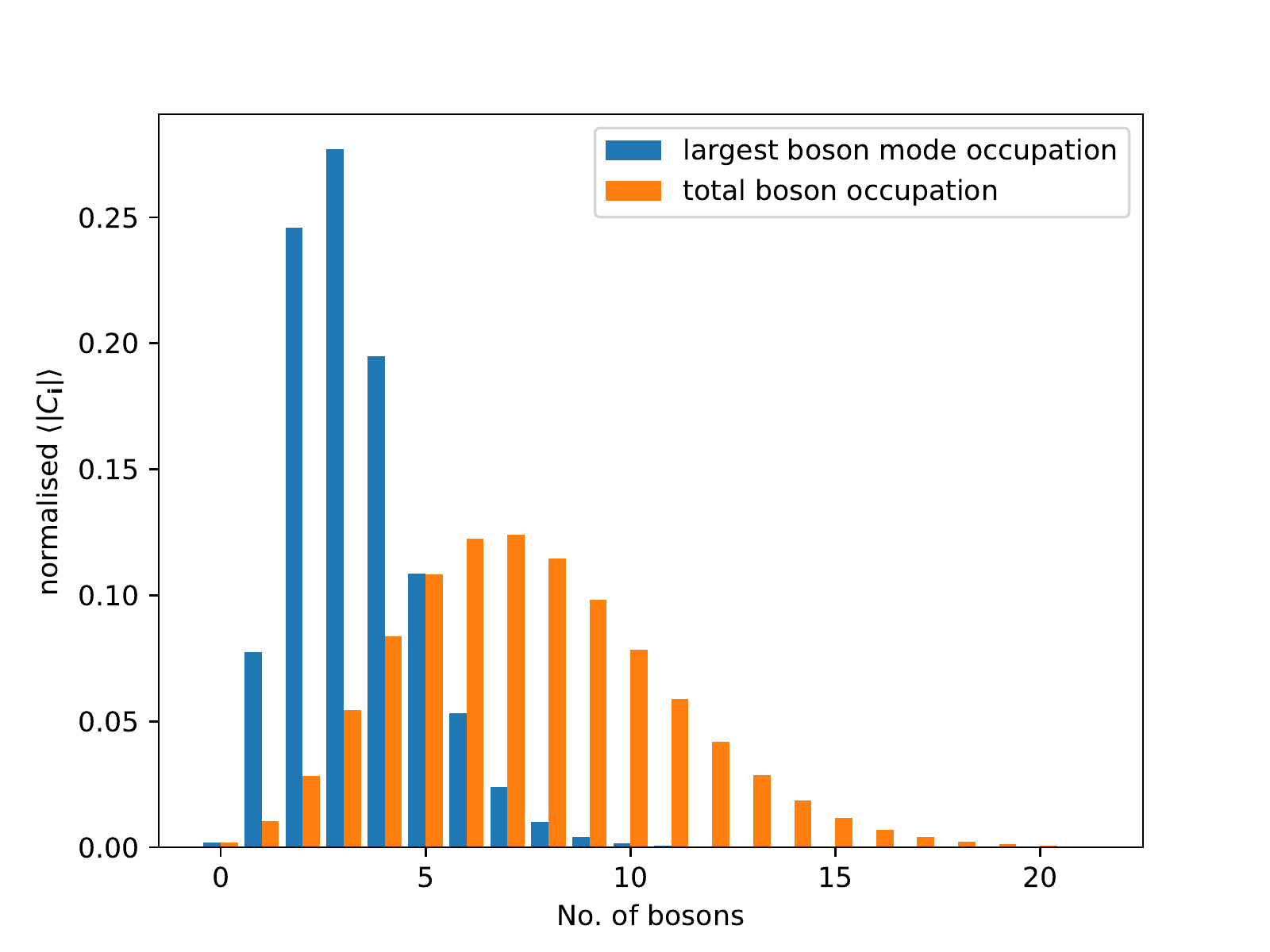}
        \caption{\vtwo{Histograms of FCIQMC walker weight according to either the largest single mode boson occupation of the configuration (blue), or the sum of the boson occupation over all bosonic modes (orange). 
        Note that appreciable amplitude of the walkers in this system reside on configurations with up to ten bosons in any one mode, reflecting the energetic impact of the bosonic truncation seen in (a).
        The calculation was performed with $n_\mathrm{boson}=255$ and a large enough $\bar{N}_W$ ($\sim 2\times 10^5$) to obtain the correct shift without the use of reweighting.}}
        \label{fig:nbos_hists}
    \end{subfigure}
    \caption{\vtwo{Demonstration of the effect of many-body wavefunction truncation due to the imposition of boson mode occupation cutoffs for a small Hubbard--Holstein chain without importance sampling ($L=8$, $n_\mathrm{elec}=8$, $U=2.0t$, $\omega_0=5.0t$, $g=\sqrt{10.0}t$).}}
    \label{fig:nboson}
\end{figure}

We now turn to a parameter regime of the 1D Hubbard--Holstein model where the bosonic renormalization effects are strong enough to induce a Peierls charge ordering, breaking the small Mott order present. This anti-adiabatic phase will be characterized by $U=2t$, $\omega_0=5.0t$, and $g=\sqrt{10.0}t$. In this regime, we expect a much slower convergence of the energy with respect to increasing boson occupation, and we can explicitly verify that this convergence correctly reproduces the infinite boson occupation limit, without any increase in bias or statistical error in the FCIQMC simulations. These results are shown in Fig.~\ref{fig:conv_nboson}, where explicit boson occupations up to eight are required for convergence to within the small FCIQMC errorbars, with the remaining population control bias accounting for $\sim0.15t$. These FCIQMC calculations were performed without importance sampling with $\sim 800$ walkers at all values of $n_\textrm{boson}$, including the infinite boson limit, which agrees with the $n_\textrm{boson}=10$ result both in value and magnitude of stochastic error. This indicates that no additional difficulties arise in the FCIQMC by fully relaxing this boson occupation constraint, which would not be possible to obtain with exact diagonalisation. \vtwo{In Fig.~\ref{fig:nbos_hists}, the fraction of wavefunction amplitude histogrammed according to either the sum of boson occupation, or the largest single mode occupation can be seen, illustrating non-negligable walker weight for this system on configurations with up to ten bosons in any single mode.} We find a best FCIQMC estimate of this point in the parameter space of the 8-site 1D Hubbard--Holstein model of $-28.612(6)t$.

\subsection{Approach to the Thermodynamic Limit}

With the ability of FCIQMC to faithfully and efficiently sample from unconstrained bosonic occupations, we now turn to increasing the physical system size, and demonstrate convergence to the thermodynamic limit of the model. We remain in the anti-adiabatic Peierls regime where strong bosonic coupling distorts the charge density and large bosonic occupations are required to be sampled as shown previously in Fig.~\ref{fig:conv_nboson}. 

Having recognised the benefits of the use of importance sampling, we return to the matter of optimising a guiding wavefunction. Figure \ref{fig:imp_samp} shows the shift energy improvement with respect to $\alpha$ for a selection of HH chains. In each of these systems the shift converges from above, and so it is valid to take the $\alpha$ responsible for the most negative energy difference to be the optimal value.

In Figure \ref{fig:imp_samp_long}, the \vtwo{absolute energy density correction due to importance sampling for a selection of chain lengths} for a constant computational cost $N_W$ is evident, with an $\alpha$ of $\sim 0.05$ conferring optimal bias reduction regardless of chain length. These data demonstrate that optimisation of the $\alpha$ parameter can be done on cheaper, low $N_W$ FCIQMC evolutions before the resulting optimal value is used with larger $N_W$ values, and in conjunction with the {\it a posteriori} reweighted shift correction. \vtwo{Neither of these measures are capable of eliminating the bias completely, but they are effective means of accelerating the systematic improvement with respect to $N_W$.}

\begin{figure}
    \centering
    \includegraphics[width=0.45\textwidth]{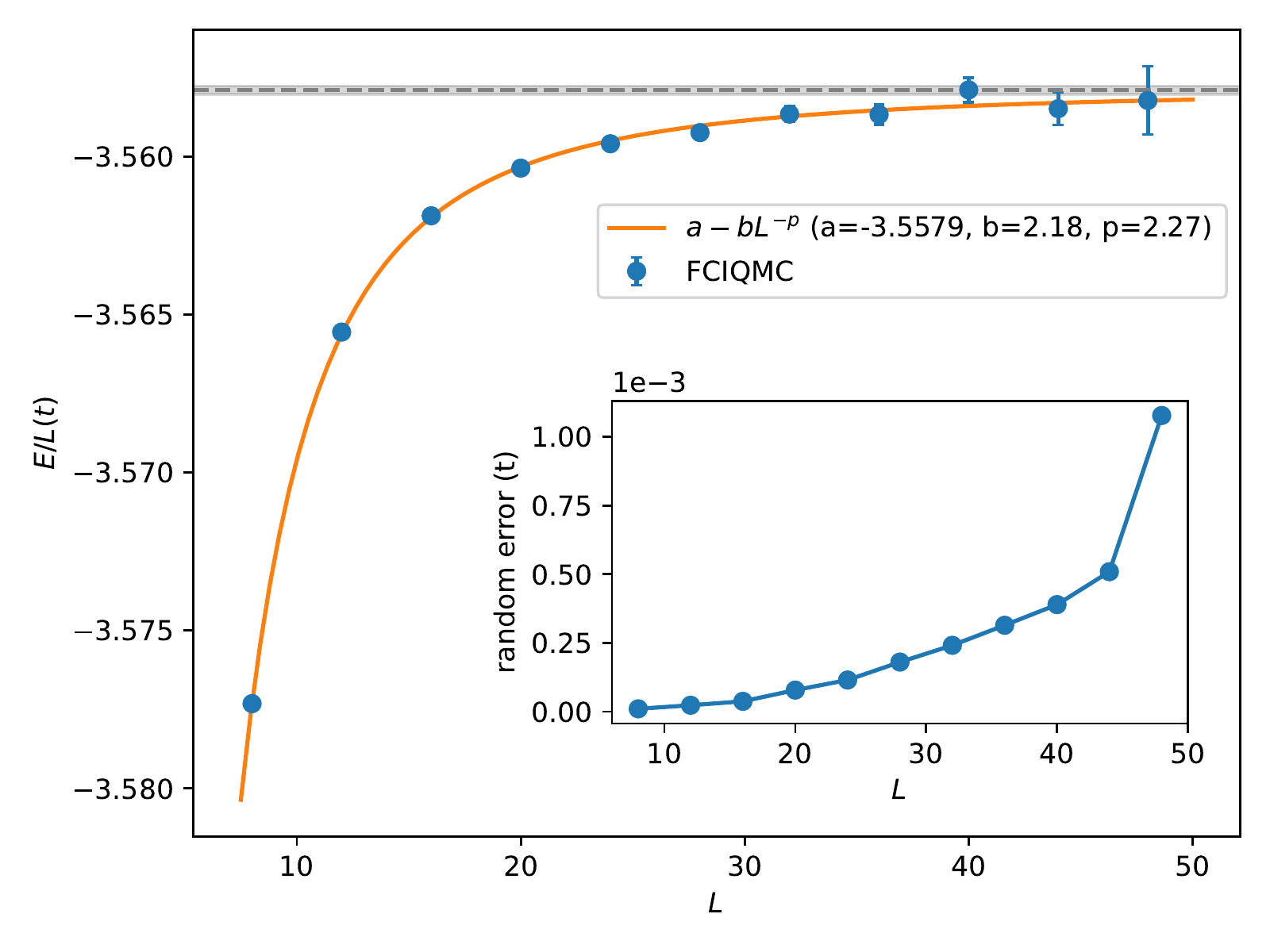}
    \caption{Convergence of the FCIQMC energy per site of the Peierls regime 1D Hubbard--Holstein model ($U=2.0t$, $\omega_0=5.0t$, $g=\sqrt{10.0}t$). Approximately $10^7$ walkers were used in each calculation, with importance sampling and the reweighting correction both applied to obtain unbiased results.
    The curve follows a power law fit to these data giving a thermodynamic limit energy density of $-3.55790(13)t$.
    The inset shows the random error in the corrected shift as a function of the chain length.
    \label{fig:conv_nsite}}
\end{figure}

Fig.~\ref{fig:conv_nsite} shows the convergence of the FCIQMC energy per site as the chain length increases to 48 sites, with $\sim 10^7$ walkers in each simulation.
A power law extrapolation of the energies achieves a highly accurate unbiased FCIQMC prediction for the thermodynamic limit energy density of $-3.55790(13)t$ per site for this point in the phase diagram.
\vtwo{The fitting was repeated with the omission of the longer chains which are associated with the largest random error; the resulting energy densities of $-3.55782(18)t$ for $L\leq42$ and $-3.55812(4)t$ for $L\leq24$ are found to agree with the $L\leq48$ estimate within standard errors. The latter result indicates a $\sim 2\times 10^{-4}t$ under-correction of the bias for $L>24$, but this is within standard error and expected given that $N_W$ was not increased with $L$.}
Also shown is the increase in the random error (as estimated by blocking analysis) of the corrected shift with respect to $L$. Since each system was tackled with an equal $N_W$ and in the same amount of CPU time, this error can be identified as directly corresponding to the scaling in the effective cost of the calculations for the unbiased energy to a given statistical uncertainty as a function of increasing chain length.

\section{FCIQMC for Electron-Boson Embedded Models} \label{sec:embeddedFCIQMC}

While a more extensive investigation of the Hubbard--Holstein model and its phase diagram with FCIQMC is planned for future work, we want to also stress the generality of the electron-boson FCIQMC approach and implementation, including to more complex interactions and couplings, and in the presence of sign problems. Rather than an {\it ab initio} electron-phonon hamiltonian, we will consider the challenge of solving an auxiliary effective electron-boson hamiltonian that arises within the context of a recently-developed quantum embedding theory. Here, the bosons in the embedded hamiltonian are fictitious, but their presence mimics the effect of interactions between an electronic fragment of a system with its (potentially large) electronic environment. In this way, the bosons describe the coupling of local electronic degrees of freedom to long-range collective electronic excitations in its environment. While these collective excitations represent composite fermionic quasiparticles, they obey (quasi-)bosonic statistics. Bosons therefore naturally arise in the description of wholly electronic systems. The hamiltonian of the embedded cluster in this context has a ``dense'' coupling of the bosons to arbitrary fermionic hopping in the fragment electronic subsystem, as well as a dense four-point electronic interaction and presence of a sign problem. Furthermore, reproducing the properties of the full system in this approach requires a sampling of reduced density matrices of this model hamiltonian, and we will investigate the fidelity by which these important quantities (including the coupled electron-boson reduced density matrix) can be sampled in FCIQMC.

This recent quantum embedding theory, which maps an electronic system to an auxiliary electron-boson coupled problem describing a small fragment of the original system, is dubbed the ``extended density matrix embedding theory'' (EDMET)\cite{PhysRevB.104.245114}, building on a ``parent'' mean-field quantum embedding formalism. The resulting local cluster hamiltonian has the form of Eq.~\ref{eq:ham}. The fermionic degrees of freedom are comprised of a small fragment of the sites of the full system, coupled to a fermionic bath of the same dimensionality. The hamiltonian in this fermionic space is algebraically constructed to ensure that the mean field properties of the fragment match between the cluster model and full system. In a similar spirit, the bosonic modes are constructed algebraically from the random phase approximation (RPA) solution over the full system, in order to require that the RPA description of the two-body fragment properties match those of the RPA over the full system. In this way, a simple construction rigorously maps from a large electronic system to a coupled electron-boson model for a fragment of the solution, with the properties of the whole system accessible from the reduced density matrices of this embedded problem. Full details of the method and the specifics of the cluster hamiltonian construction can be found in Ref.~\onlinecite{PhysRevB.104.245114}.

We now consider the ability of FCIQMC to provide the required accuracy in the solution of this embedded cluster hamiltonian. Two differences from the form of Eq.~\ref{eq:ham} exist in the resulting cluster model; $g_n=0$ (i.e. there are no single boson terms not coupled to the charge density), and 
the electron-boson coupling term is zero when the fermionic indicies $i$ and $j$ co-incide, if the fermionic cluster basis is canonicalized to a mean-field representation (i.e. the bosons represent couplings to the fragment density fluctuations, rather than the mean-field density, and so ZPMR is not effective). To date, only exact diagonalization has been used as a solver of these embedded hamiltonians, which puts a severe constraint on both the size of the fragment spaces which can be chosen, as well as the boson occupation truncation to ensure computational tractability. Physically, this truncation corresponds to limiting the energy-scales of long-range plasmonic excitations that the fragment can couple to, and it would be advantageous to not truncate these effects, and so alternative solvers for these auxiliary model hamiltonians is urgently sought. Using electron-boson FCIQMC will however require an ability to sample the reduced density matrices (RDM) of the coupled electron-boson model in FCIQMC, including electron-boson terms (which will also be useful for other property estimates in a more general context).



\subsection{Reduced Density Matrix Sampling}

The standard way to stochastically estimate (off-diagonal) elements of RDMs in FCIQMC is to repurpose the spawning step of the algorithm. This step can be considered the act of randomly drawing a configuration pair, and a product of two instantaneous walker occupations taken from statistically independent replica simulations, which can then be used to estimate their contribution to unbiased RDMs \cite{doi:10.1063/1.4986963, doi:10.1063/1.4904313}.
This has proven a successful means of sampling one- and two-body RDMs in fermionic systems for a diversity of applications, e.g. molecular properties \cite{doi:10.1063/1.4927594, doi:10.1021/acs.jctc.8b00454}, MCSCF orbital optimisations \cite{Thomas2015_2, doi:10.1021/acs.jctc.5b00917, doi:10.1021/acs.jctc.5b01190, doi:10.1021/acs.jctc.1c00936} and contributions from internally contracted perturbers in multireference perturbation theories \cite{doi:10.1063/1.5140086, doi:10.1080/00268976.2020.1802072}. However, this approach makes an approximation that the set of configurational pairs, $(\mathbf{i}, \mathbf{j})$ with non-zero coefficient product which contribute to the RDMs of interest ($\Gamma$-connections) is equal to the set of $(\mathbf{i}, \mathbf{j})$ pairs connected by a non-zero Hamiltonian matrix element ($H$-connections).
This is required as \vtwo{it is impractical} to essentially take the outer product of a sparse, computationally distributed representation of the FCI wavefunction so that all $\Gamma$-connections are explicitly included, and so the feasibility of accurate, fully stochastic RDM estimates within an FCIQMC computational framework is predicated on the success of the assumed equivalence of $H$- and $\Gamma$-connections. 

A stark example of the breakdown of this assumption is in the case of Hartree--Fock canonical orbital representations, in which single excitations of the Hartree--Fock determinant can have substantial coefficients, but do not interact with the reference through the hamiltonian, $H$, so relying on spawning alone to sample the 1RDM would neglect some large contributions. This class of $\Gamma$-connections is therefore explicitly included as a separate contribution.
An approach to including these contributions more widely, is to ensure that important $\Gamma-$connected configuration pairs (which may not be significantly weighted $H-$connected configurations) are included in a ``deterministic'' space, which is already widely used in FCIQMC propagation\cite{doi:10.1063/1.4920975, PhysRevLett.109.230201}. Here, coefficients on configurations designated to comprise the deterministic subspace are collected from across all processes and multiplied exactly by a sparse representation of the Hamiltonian projected into that subspace.
In a semi-stochastic RDM calculation, another sparse matrix object can be initialised which only stores those $\Gamma$-connections which are not also $H-$connections so that {\it all} RDM contributions between configurations in the deterministic subspace are made without approximation. As the size of the deterministic subspace increases, any error of the kind described above is rigorously removed. 

However, ensuring all $\Gamma-$connected configurations are included in a deterministic subspace may not always be efficient or practical. Sampling of the remaining $\Gamma$-connections (those which do not involve pairs of configurations in the deterministic subspace) can be achieved by the implementation of special ``ghost'' excitation generators, which are propagated separately to the spawning events of FCIQMC and only carry the instantaneous population of the parent configuration (ignoring generated connections that would already be accounted for by the deterministic subspace), specifically for RDM accumulation. Such a situation has already been covered in detail in the accumulation of the three-body RDM and four-body contracted auxiliary matrices for MRPT2\cite{doi:10.1063/1.5140086}, where these higher-rank contributions are explicitly outside the rank of the standard electronic hamiltonian.
The necessity to confront the discrepancy between $H-$ and $\Gamma-$connected configuration pairs manifests again in the present context, since the fermion-boson RDM $\langle \hat{c}^\dagger_{i}\hat{c}_{j}\hat{a}^\dagger_n \rangle$ would not receive contributions to any elements of the form
$\langle \hat{c}^\dagger_{i}\hat{c}_{i}\hat{a}^\dagger_n \rangle$ since there are no direct interactions ($H-$connections) between the mean-field electronic densities and boson (de-)excitations. 
Thus, a whole class of configuration pairs is omitted by assuming $H-$ and $\Gamma$-connection equivalence when propagating the EDMET cluster Hamiltonian. The same would be true not just for the EDMET hamiltonian, but also general hamiltonians where the ZPMR transformation of the bosons has been applied (generalized to remove coupling to a mean-field density).

A possible solution is to implement a uniform excitation generator to draw ``ghost'' connections of the form $|\mathbf{i}\rangle \rightarrow \hat{a}^\dagger_n|\mathbf{i}\rangle$, $|\mathbf{i}\rangle \rightarrow \hat{a}_n|\mathbf{i}\rangle$, purely for the purpose of sampling these neglected contributions to the fermion-boson coupled RDM, as was done for triple excitations in the generation of 3-body RDMs. However in the present work, this class of $\Gamma$-connection was taken into account by the deterministic space alone, and so any RDM sampling must be checked for convergence with respect to the number of configurations in this subspace, denoted $N_\text{SS}$. It should be stressed that a more general hamiltonian which also coupled to the full quadratic fermion terms would not encounter this problem.

\subsection{Extended Hubbard Model}

We consider FCIQMC as the auxiliary hamiltonian ``solver'' within an EDMET framework, applied to the 1D Hubbard model with extended interactions. This purely fermionic model can be written as
\begin{equation}
    {\hat H}_\mathrm{Ex-Hubbard} = {\hat H}_\mathrm{Hubbard} + V \sum_{\langle i,j \rangle} \hat{n}_{i} \hat{n}_{j} ,
\end{equation}
and we aim to converge to the thermodynamic limit of the model. The additional nearest neighbour repulsion fundamentally changes the phase diagram from the simple Mott insulating phase of the 1D ${\hat H}_\textrm{Hubbard}$, to one where the competing $V$ term can drive the system into a charge ordered phase to minimize nearest neighbour density-density repulsion. The phase diagram therefore roughly consists of Mott insulating phases for $V \lesssim U/2$ and a charge ordered phase for $V \gtrsim U/2$ \cite{PhysRevB.65.155113, PhysRevLett.88.056402, PhysRevLett.89.236401, PhysRevLett.99.216403}. Rather than solve this directly in the large system limit, we aim to solve a local model of a cluster of $n_\textrm{frag}$ local sites, coupled to fermionic and bosonic degrees of freedom approximating the coupling and entanglement with the wider system (a bath space). The long range charge fluctuations induced by the nearest neighbour interactions will then be modelled by the random phase approximation, and used to map to a local model of bosons coupled to the interacting local fermionic space. Charge ordering in the model can then be induced by these bosons in the auxiliary model, rather than collective long-range fermionic fluctuations. The energy density of the model over the fragment space can then be computed from the one- and two-body fermionic density matrices of the model, as well as the boson-fermion coupled density matrix, as $\langle \hat{c}_i^{\dagger} {\hat c}_j \hat{a}_n \rangle$ and $\langle \hat{c}_i^{\dagger} {\hat c}_j \hat{a}^{\dagger}_n \rangle$. 

\begin{figure}
    \centering
    \includegraphics[width=0.45\textwidth]{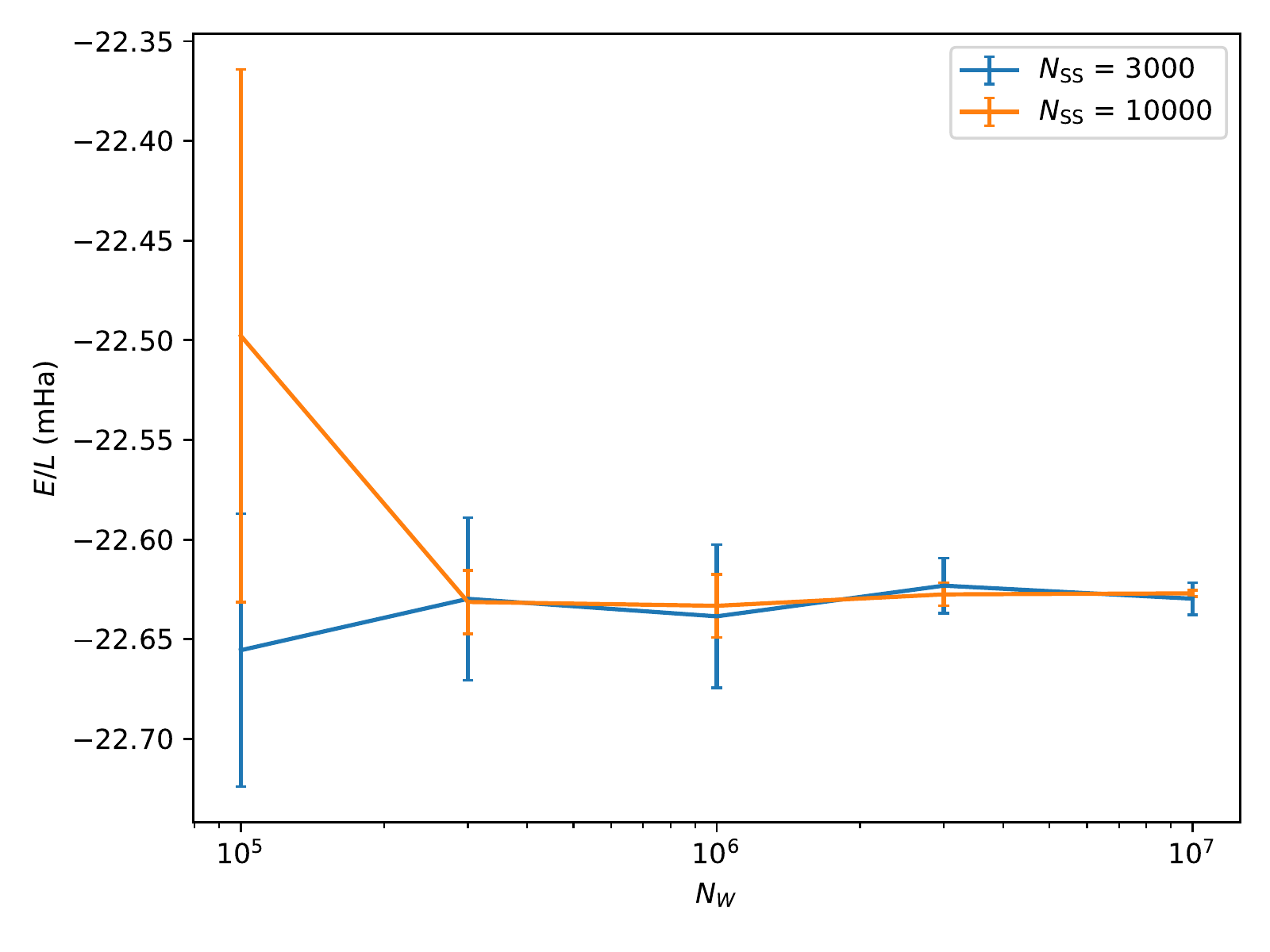}
    \caption{The FCIQMC-EDMET energy per site for an extended Hubbard model ($L=42, U=2.0, V=1.0$) as determined from multiple independent FCIQMC estimations of the required RDMs of a 4-site fragment size. This leads to a auxiliary hamiltonian consisting of 8 fermions and 10 bosons, with no cutoff on bosonic occupation. The calculations were carried out for two different sizes of semi-stochastic subspace to ensure convergence with respect to RDM contributions not generated through spawning
    \label{fig:edmet_nw}}
\end{figure}

We consider a specific point in the phase diagram with $U=2t$ and $V=1t$, which is very close to the phase transition point between Mott insulator and charge ordered phases \cite{PhysRevLett.99.216403}. We expect therefore that the model will feature significant long range two-point fluctuations, which are hard to model in a local model and will have to be mimicked via strong coupling to the bosonic bath. We first consider a four-site fragment space, to ensure convergence of the system properties with FCIQMC. 
Mapping a 42 site chain (close to the thermodynamic limit for the energy density) to the auxiliary hamiltonian of Eq.~\ref{eq:ham}, with eight fermionic degrees of freedom and 10 boson modes, we consider the convergence with number of walkers in Fig.~\ref{fig:edmet_nw}. Additionally, we check that changes to the deterministic space does not change the fidelity of the RDMs, from which the energies are computed. 

We find a very rapid convergence of the FCIQMC energies with respect to increasing walker number, with all values between $N_W=10^5$ and $N_W=10^7$ statistically indistinguishable within their random error bars, estimated from multiple calculations with different random number seeds. Convergence with walker number is expected due to the use of the ``initiator'' approximation in the FCIQMC dynamics to control the sign problem. In these simulations, the initiator approximation is employed with the threshold walker number of $n_a=3$ in order to define an initiator configuration. This is without any restriction of bosonic occupation, and shows that even in the presence of a sign problem in this model, these systems can be faithfully converged, relying on a combination of annihilation events and the initiator approximation to converge. Furthermore, changing the size of the deterministic space ($N_\mathrm{SS}$) does not further change the quality of the results, but does reduce the statistical fluctuations. The scale of these fluctuations is however already very small, with the EDMET energy density for this fragment size reliably found within $10^{-4}t$ by $\sim 10^6$ walkers, demonstrating the applicability of fermion-boson FCIQMC as a cluster solver for this embedding method.

\begin{figure}
    \centering
    \includegraphics[width=0.45\textwidth]{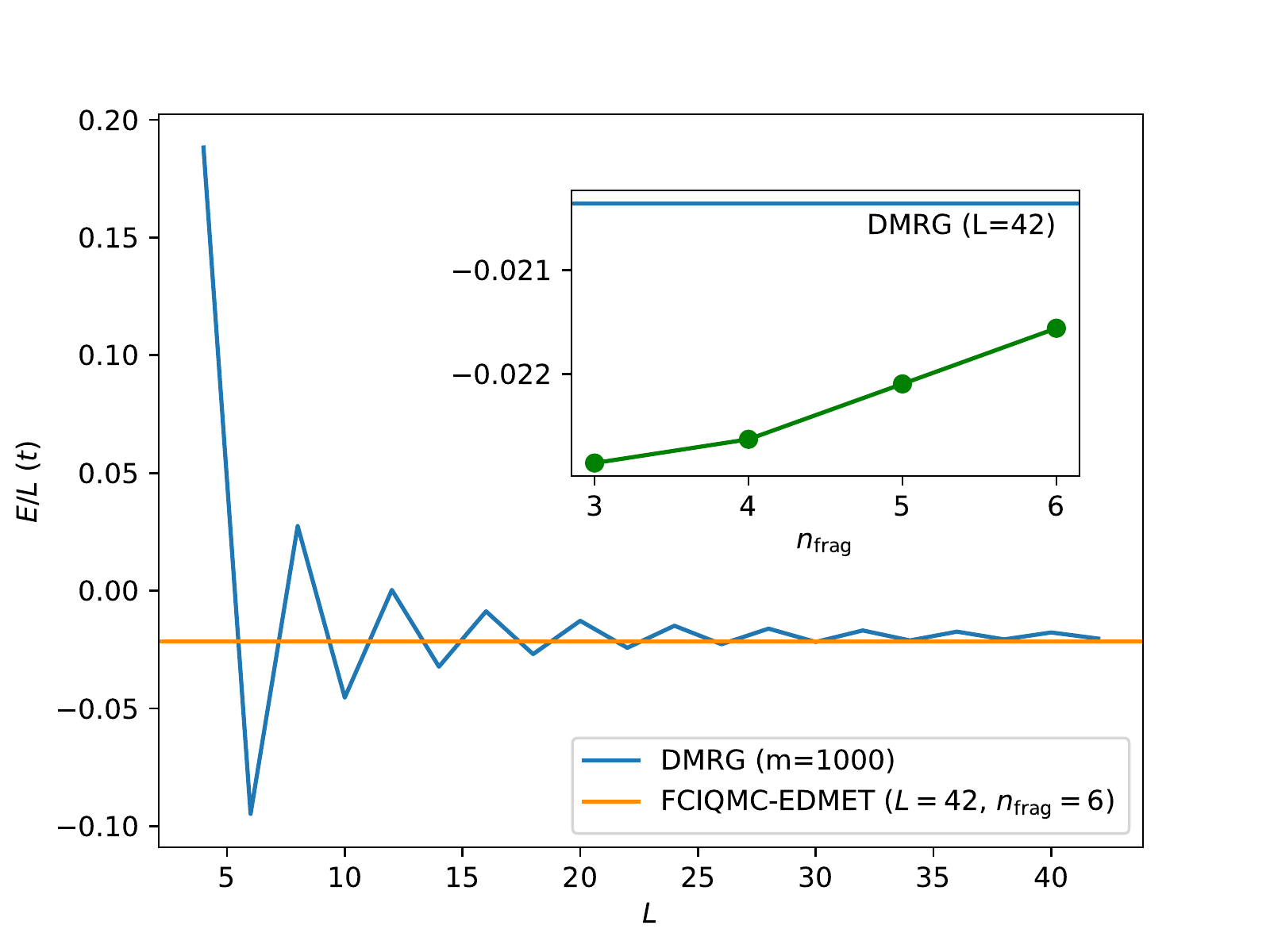}
    \caption{Comparison of DMRG and EDMET using the FCIQMC cluster solver for a half-filled extended Hubbard model ($L=42, U=2.0, V=1.0$) as the system size increases. The inset shows the systematic improvement of the EDMET per-site energy with respect to the size of the number of sites in the fragment. Convergence with fragment size is imperceptible on the scale of the main plot.}
    \label{fig:edmet_nimp_dmrg}
\end{figure}

We then consider the convergence of the FCIQMC-EDMET energies as the fragment size increases. This model can be effectively solved within the density matrix renormalization group (DMRG) given the one-dimensional nature of the model, for which we use the {\tt block} code, with 1000 renormalized states retained in the matrix product state of each site. In Fig.~\ref{fig:edmet_nimp_dmrg} we show the convergence of the energy density to the thermodynamic limit for this parameter regime as the chain length increases. Taking the largest chain length of 42 sites, we also map to the local coupled electron-boson cluster model with EDMET for different numbers of fragment sites, and solve with FCIQMC. The convergence with $n_\mathrm{frag}$ is shown in the inset, and demonstrates the monotonic convergence, to an accuracy in the ground state energy of $\sim 10^{-3}t$, with both the EDMET mapping and the FCIQMC solver giving systematic errors smaller than the residual finite size errors. The $n_\textrm{frag}=6$ cluster problem consists of a 12 orbital, 12 electron fermionic space, coupled to 21 auxiliary bosonic modes, which is easily treated, converging within $\sim 10^{-4}t$ between $10^6$ and $3\times 10^6$ walkers after 3000 RDM accumulation cycles.

Further improvements in the FCIQMC-EDMET approach can be found via self-consistency, which we neglect in this demonstration. In this, the RDMs are used to update both the mean-field and interaction kernel of the RPA, subsequently modifying the resulting cluster hamiltonian in the bath space. We will explore the potential of FCIQMC-EDMET in future work, with the main conclusion here being the applicability of FCIQMC to electron-boson systems for more general hamiltonians and as a solver in this workflow, with the initiator approximation and annihilation events of the sparse representation allowing for faithful convergence of both energies and RDMs even in the presence of a sign problem in these infinite Hilbert spaces.

\section{Conclusions and Outlook}

We have developed an extension of the FCIQMC method to enable application to coupled electron-boson systems, avoiding the truncation in bosonic occupancy which can limit the efficiency of other wave function based approaches. We describe the changes for efficient excitation generation in these systems, as well as considerations in sampling reduced density matrices. We demonstrate the convergence and applicability of the method in two very different physical contexts. Firstly it was applied to a sign-problem-free system (the 1D Hubbard--Holstein model) where sparse sampling and care to control for population bias was shown to enable unbiased results in the thermodynamic limit. Furthermore, the method was used as a solver of an embedded hamiltonian within the extended density matrix embedding method, which maps large electronic systems to local interacting electron-boson auxiliary models to solve. These generalized electron-boson models induce sign problems in the FCIQMC dynamics, which we show can be easily controlled via standard FCIQMC techniques, without having to resort to truncation of the boson occupancy. Furthermore, the extraction of the full system energy density requires a faithful sampling of the reduced density matrices (including the electron-boson RDM), which we demonstrate are faithfully sampled and compare to DMRG reference values in the full model.

\vtwo{Although the results presented here are for the ground state, FCIQMC's dynamical adaptations are in principle completely compatible with mixed fermion-boson Hamiltonians. In this regard however, it should be noted that immunity to the sign problem --- which is key to the method's efficacy in the Hubbard--Holstein model --- is a state-specific attribute, and that any attempt to sample excited states within the walker dynamic, either by explicit projection \cite{Blunt2017} or real-time propagation \cite{guther2018} will necessitate annihilations, thereby reintroducing a sign problem. Krylov projection \cite{Blunt2015_2, Blunt2018} on the other hand, does not interfere with the sign-problem-free propagation towards the ground state; but attempting to sample matrix elements, particularly those of the overlap matrix between Krylov space-spanning replica walker populations would be marred by the highly delocalised nature of the wavefunction in the site basis. None of these caveats however are especially applicable to the more localised and sign-problematic low-lying eigenfunctions encountered in the systems described in Section~\ref{sec:embeddedFCIQMC}.
}

Having established the applicability of FCIQMC in \vtwo{a new domain}, a wide array of physical phenomena are now available for study, and future work must apply the approach to open problems to assess the merits and limits of the method in the context of complementary methods. We expect particular efficiency of FCIQMC in cases of general interacting hamiltonian forms, with many bosons that are relatively weakly coupled. This case allows for a sparse representation of the wave function to be efficiently sampled within FCIQMC, and enables the annihilation and initiator approximations to work at their best, without requiring {\it a priori} truncations on bosonic occupation. The use of FCIQMC as a solver within EDMET is an example of this type of setting, and will be further explored, including self-consistency in future work, where the ability to obtain unbiased RDMs is a particular strength of the method compared to more traditional QMC approaches to electron-boson problems.

\acknowledgements

G.H.B. gratefully acknowledges support from the Royal Society via a University Research Fellowship, as well as funding from the European Union's Horizon 2020 research and innovation programme under grant agreement No. 759063. We are grateful to the UK Materials and Molecular Modelling Hub for computational resources, which is partially funded by EPSRC (EP/P020194/1).

%

\appendix
\section{Zero Phonon Mode Removal}
\label{app:zpmr}
When studying the Hubbard--Holstein model with periodic boundary conditions, it is routine to bring about a reduction in the coupling between the electron and boson sector by applying the zero phonon mode removal (ZPMR) transformation to the bosonic second quantised operators. This can be written as
\begin{equation}
    \hat{a}_m\rightarrow\hat{a}_m-\frac{g}{\omega_0}\langle n \rangle
\end{equation}
where $\langle n \rangle = N_\mathrm{elec}/L$ stands for the average electronic density per site,
Under this transformation, the electron-boson coupling becomes
\begin{equation}
    g\hat{n}_m \left(\hat{a}_m^\dagger + \hat{a}_m - \frac{2g}{\omega_0}\langle n \rangle\right) .
\end{equation}
Since the fermion number is fixed, the average fermion occupation per site is known by symmetry, and the resultant on-site one-electron contribution amounts to an energy shift of $-2g^2 N_\mathrm{elec}{}^2 / (\omega_0 L)$. Likewise, the boson energy contributions transform to
\begin{equation}
    \omega_0\left(\hat{a}_m^\dagger\hat{a}_m  - \frac{g}{\omega_0}\langle n \rangle (\hat{a}_m^\dagger + \hat{a}_m) + \frac{g^2}{\omega_0^2}\langle n \rangle^2\right)
\end{equation}
from which the second term shows that there are now single boson (de-)excitations that are uncoupled to the electronic modes, and an additional constant. After summation over sites, this gives another constant energy shift of $g^2 N_\mathrm{elec}{}^2 / (\omega_0 L)$, resulting in an overall energy shift of $-g^2 N_\mathrm{elec}{}^2 / (\omega_0 L)$ to the ZPMR Hamiltonian spectrum relative to the untransformed definition.

This transformation redefines the vacuum state in the boson modes in such a way that the boson occupation-truncated wave function converges faster to the untruncated wave function with respect to $n_\mathrm{boson}$, making larger systems tractable by exact eigensolvers provided that the coupling to the bosons is weak. Physically, this amounts to decoupling the bosons from the static charge density of the system, ensuring only coupling to the instantaneous quantum density fluctuations (higher moments of the density distribution) remain.
It is tempting to bring the ZPMR transformation into effect in the hope of improving the performance of FCIQMC by condensing the walker distribution around configurations with lower-occupied modes. Indeed, this is likely to be advantageous for general hamiltonians. However, for the Hubbard--Holstein model, it must be noted that the density-uncoupled (de-)excitations between boson number sectors would always carry the opposite sign to the density-coupled (de-)excitations. This introduces a sign problem, which would negate the advantages in wave function compactification achieved by the transformation.
Therefore, use of the ZPMR transformation in the present work is restricted to obtaining faster converging FCI energies with respect to boson occupation cutoff for small \vtwo{systems}, to allow comparison to FCIQMC results.

\end{document}